\documentclass[fleqn,12pt,twoside]{article}
\usepackage[headings]{espcrc1}

\readRCS
$Id: espcrc1.tex,v 1.2 2004/02/24 11:22:11 spepping Exp $
\usepackage{graphicx}
\usepackage[figuresright]{rotating}

\newcommand {\bea}{\begin{eqnarray}}
\newcommand {\eea}{\end{eqnarray}}
\newcommand {\be}{\begin{equation}}
\newcommand {\ee}{\end{equation}}

\newcommand{\AmS}{{\protect\the\textfont2
  A\kern-.1667em\lower.5ex\hbox{M}\kern-.125emS}}

\title{Non-Fermi Liquid Effective Field Theory of Dense QCD Matter}

\author{Thomas Sch\"afer\address{Department of Physics,  
        North Carolina State University, \\ 
        Raleigh, NC 27695}}
       
\runtitle{Non-Fermi Liquid Effective Field Theory}
\runauthor{T.~Sch\"afer}

\begin{document}

\maketitle

\begin{abstract}
We review an effective field theory for the non-Fermi liquid
regime of dense QCD matter. Non-Fermi liquid effects arise
due the presence of unscreened magnetic gluon exchanges. We
show that there is a systematic low energy expansion in 
fractional powers and logarithms of energy. We discuss the 
validity of some standard theorems of Fermi liquid theory.
\end{abstract}

\section{Non-Fermi liquid effective field theory}
\label{sec_nfl}

 At high baryon density the relevant degrees of freedom are 
particle and hole excitations which move with the Fermi 
velocity $v$. Since the momentum $p\sim v\mu$ is large, 
typical soft scatterings cannot change the momentum by very 
much and the velocity is approximately conserved. An effective 
field theory of particles and holes in QCD is given by 
\cite{Hong:2000tn,Schafer:2005mc}
\be
\label{l_hdet}
{\cal L} =\psi_{v}^\dagger \left(iv\cdot D
   - \frac{1}{2p_F}D_\perp^2 \right) \psi_{v}
   + {\cal L}_{4f}
   + {\cal L}_{HDL} 
   -\frac{1}{4}G^a_{\mu\nu} G^a_{\mu\nu}+ \ldots ,
\ee
where $v_\mu=(1,\vec{v})$. The field $\psi_v$ describes particles 
and holes with momenta $p=\mu(0,\vec{v})+k$, where $k\ll\mu$. We will 
write $k=k_0+k_{\|}+k_\perp$ with $\vec{k}_{\|}=\vec{v}(\vec{k}\cdot 
\vec{v})$ and $\vec{k}_\perp = \vec{k}-\vec{k}_{\|}$. ${\cal L}_{4f}$ 
denotes four-fermion operators in the BCS and zero sound channel. At 
energies below the screening scale $g\mu$ hard dense loops have to be 
resummed. The generating functional for hard dense loops in gluon 
$n$-point functions is given by \cite{Braaten:1991gm}
\be 
\label{S_hdl}
{\cal L}_{HDL} = -\frac{m^2}{2}\int\frac{d\hat{v}}{4\pi} 
  \,G^a_{\mu \alpha} \frac{v^\alpha v^\beta}{(v\cdot D)^2} 
G^b_{\mu\beta},
\ee
where $m^2=N_f g^2\mu^2/(4\pi^2)$ is the dynamical gluon mass and 
$\hat{v}$ is a unit vector in the direction of $\vec{v}$. 

 The hard dense loop action describes static screening of electric 
fields and dynamic screening of magnetic modes. Since there is no 
screening of static magnetic fields low energy gluon exchanges are 
dominated by magnetic modes. The resummed transverse gauge boson 
propagator is given by
\be
\label{d_trans}
D_{ij}(k) = \frac{\delta_{ij}-\hat{k}_i\hat{k}_j}{k_0^2-\vec{k}^2+
i\eta k_0/|\vec{k}|} ,
\ee
where $\eta=\frac{\pi}{2}m^2$ and we have assumed that $|k_0|<|\vec{k}|$. 
We observe that the gluon propagator becomes large in the regime 
$|\vec{k}|\sim (\eta k_0)^{1/3}\gg k_0$. This leads to an unusual scaling 
behavior of Green functions in the low energy limit. Consider a generic 
Feynman diagram and scale all energies by a factor $s$. Because of the 
functional form of the gluon propagator in the Landau damped regime 
gluon momenta scale as $|\vec{k}|\sim s^{1/3}$. This implies that the 
gluon momenta are much larger than the gluon energies. The quark dispersion 
relation is $k_0\simeq k_{||}+k_\perp^2/(2p_F)$. The only way a quark can 
emit a very spacelike gluon and remain close to the Fermi surface is if 
the gluon momentum is transverse to the Fermi velocity. We find
\be 
k_0 \sim s, \hspace{0.5cm}
k_{||}\sim s^{2/3},\hspace{0.5cm}
k_\perp \sim s^{1/3},
\ee
and $k_0\ll k_{||}\ll k_\perp$. In the low energy regime propagators 
and vertices can be simplified even further. The quark and gluon 
propagators are 
\be
 S^{\alpha\beta}(p) = \frac{i\delta_{\alpha\beta}}
       {p_0 -  p_{||} - \frac{p_\perp^2}{2\mu}
              +i\epsilon \,{\mathrm{sgn}}(p_0)},
\hspace{0.75cm} 
 D_{ij}(k)  =  \frac{-i\delta_{ij}}
       {k_\perp^2-i\frac{\pi}{2}m^2\frac{k_0}{k_\perp}} \, , 
\ee
and the quark gluon vertex is $gv_i(\lambda^a/2)$. Higher order 
corrections can be found by expanding the quark and gluon propagators 
as well as the HDL vertices in powers of the small parameter $\epsilon\!
\equiv\!\omega/m$.

\begin{figure}
\includegraphics[width=5.25cm]{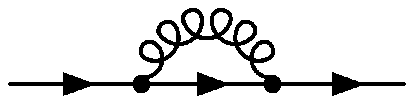}\hspace{-0.75cm}
\includegraphics[width=5.25cm]{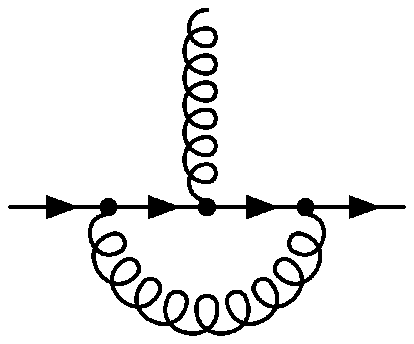}\hspace{-0.75cm}
\includegraphics[width=5.25cm]{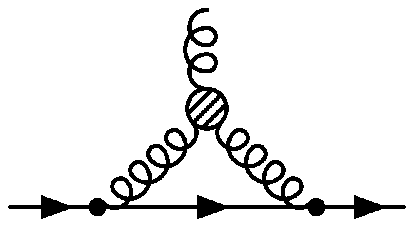}
\caption{One-loop contributions to the quark self energy 
and the quark-gluon vertex. In the magnetic regime  
the graphs scale as $\omega\log(\omega)$, $\omega^{1/3}$
and $\omega^{2/3}$, respectively. }
\label{fig_mag}
\end{figure}

 We can show that the power of $\epsilon$ 
associated with a Feynman diagram always increases with the number of 
loops and the number of higher-order vertices. One way to see this is 
to rescale the fields in the effective lagrangian so that the kinetic 
terms are scale invariant under the transformation $(x_0,x_{||},x_\perp)
\to (\epsilon^{-1}x_0,\epsilon^{-2/3} x_{||},\epsilon^{-1/3}x_{\perp})$. 
The scaling behavior of the fields is $\psi\to \epsilon^{5/6}\psi$ and 
$A_i  \to \epsilon^{5/6} A_i$. We find that the scaling dimension of 
all interaction terms is positive. The quark gluon vertex scales as 
$\epsilon^{1/6}$, the HDL three gluon vertex scales as $\epsilon^{1/2}$, 
and the four gluon vertex scales as $\epsilon$. Since higher order 
diagrams involve at least one pair of quark gluon vertices the expansion 
involves positive powers of $\epsilon^{1/3}$ and the low energy regime is 
completely perturbative.

\begin{figure}
\includegraphics[width=5cm]{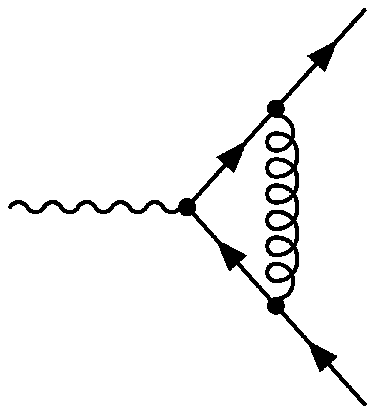}
\hspace*{1.5cm}
\includegraphics[width=5cm]{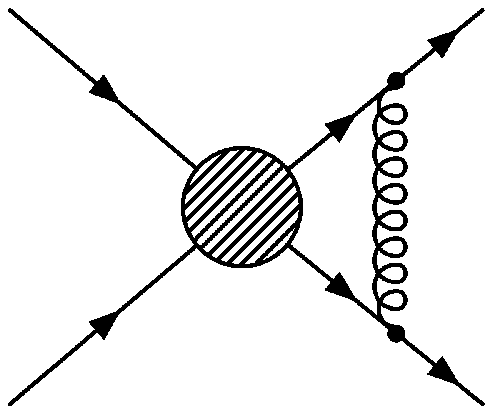}
\caption{One-loop correction to the vertex of an external 
current and the BCS interaction. Both diagrams are kinematically
enhanced and scale as $\log(\omega)$ and $\log^2(\omega)$, respectively. }
\label{fig_enh}
\end{figure}

\section{Migdal Theorem}
\label{sec_mig}

 As a simple application consider the one-loop correction to the 
quark-gluon vertex for a gluon in the non-Fermi liquid regime, see
Fig.~\ref{fig_mag}. According to the scaling rules discussed in 
the last section this correction is dominated by the Abelian 
diagram, and scales as 
\be 
\Gamma^a_\mu = gv_\mu (\lambda^a/2) \left(1+O(\epsilon^{1/3})\right).
\ee
This is a QCD version of Migdal's theorem, which states that 
the renormalization of the electron-phonon vertex is suppressed
by the ratio $\sqrt{m/M}$, where $m$ is the mass of the electron 
and $M$ is the mass of the ions \cite{Migdal:1958}. This factor is 
analogous to the small paramater in the QCD case, because it controls 
the ratio of the phonon velocity to the Fermi velocity of the 
electrons, and therefore also the ratio of the typical electron
momentum to the typical phonon momentum. 

 We note that it is essential the the gluons are spacelike. 
Consider the vertex of an external gauge field with coupling $e$.
The one-loop correction in the regime of small time-like
momenta is \cite{Brown:2000eh}
\be
 \Gamma_\mu (p_1,p_2) = \frac{eg^2}{9\pi^2} v_\mu
 \log\left(\frac{\Lambda}{\omega}\right) ,
\ee
where $p_{1,2}$ are the momenta of the two quarks, $\omega=(p^0_1+p^0_2)$ 
and the scale $\Lambda$ was determined in \cite{Schafer:2005mc}. 
In this regime higher order corrections are large and have to be 
resummed. This is exactly what also happens in the BCS and zero-sound 
channels: Propagators are kinematically enhanced, and all two-body 
ladders have to resummed. Non-planar diagrams, or Green functions 
with more legs are perturbative and follow the scaling rules discussed 
in the previous section \cite{Schafer:2005mc}.

\section{Kohn-Luttinger Theorem}
\label{sec_kl}

 One of the applications of the non-Fermi liquid effective theory is 
the calculation of higher order correction to the superfluid gap. The 
theory can also be used to study the possibility of exotic kinds
of superfluidity. Kohn and Luttinger showed that for ordinary 
Fermi liquids superfluidity takes place even if the basic 
particle-particle interaction is repulsive in all channels
\cite{Kohn:1965}. We recently studied whether this mechanism 
also operates in gauge theories and leads to superfluidity in a 
cold electron gas \cite{Schafer:2006ue}. The gap equation is 
\be
 \Delta_0 = -\frac{e^2}{8\pi^2} 
  \int \frac{d\omega\,\Delta(\omega)}{\sqrt{\omega^2+\Delta(\omega)^2}}
 \ f_l(\omega)
\ee
where $f_l(\omega)$ is the scattering amplitude of a $(p_F,-p_F)$ 
electron pair with energy $\omega$ and angular momentum $l$. We
find that for any fixed $\omega$ the amplitude $f_l(\omega)$ becomes
attractive at large $l$. This is the Kohn-Luttinger effect. However, 
for fixed $l$ the gauge theory amplitude is always repulsive at 
very small $\omega$ and superconductivity does not take place. 

\section{Luttinger Theorem}
\label{sec_l}

 Luttinger showed \cite{Luttinger:1960} that in a Fermi liquid 
the relationship between the density and the Fermi momentum is 
given by the free Fermi gas result 
\be
 \frac{N}{V} = g_d \int \frac{d^3p}{(2\pi)^3}\;\Theta (p_F-p)
\ee
even if the system is interacting. Here, $g_d$ is the degeneracy and 
$p_F$ can be defined as the momentum at which the inverse propagator 
$S^{-1}(\omega\!=\!0,p)$ changes sign. The derivation of this result 
is discussed in standard textbooks on many body theory \cite{Abrikosov:1963}. 
The main step is
\be
\frac{N}{V}=ig_d\int \frac{d^3p}{(2\pi)^3}
 \int_{-\infty}^0 \frac{d\omega}{2\pi}\frac{\partial}{\partial\omega}
 \log\left(\frac{S(\omega,p)}{S_{ret}(\omega,p)}\right)
 = -\frac{g_d}{\pi}\int \frac{d^3p}{(2\pi)^3}
\left[ \varphi(0,p)-\varphi(-\infty,p)\right]
\ee
where $S_{ret}$ is the retarded propagator and $\varphi$ is its
complex phase. Luttinger's theorem follows if the phase
goes from 0 to $\pi$ as $\omega$ goes from 0 to $-\infty$ for the 
occupied states, and remains zero for unoccupied states. In a 
gauge theory the self energy has a cut at $\omega=0$, $\Sigma(\omega) 
= c_1 \omega\log(\omega)+ c_2 \omega^{4/3} + O(\omega^{5/3})$. As
a result there is no Fermi surface in the ordinary sense as the Fermi 
velocity and the quasi-particle renormalization factor go to zero as 
$\omega\to 0$. However, Luttinger's theorem remains valid because the 
fermion propagator changes sign at $p=p_F$ and it acquires a phase 
$\varphi=\pi$ for $p<p_F$ \cite{Dzyaloshinskii:2003}. Luttinger's 
theorem is not directly applicable in the superfluid phase. 

Acknowledgments: This work was supported by US DOE
grant DE-FG02-03ER41260.

\end{document}